\documentclass[aps,prl,twocolumn,superscriptaddress,floatfix,longbibliography]{revtex4-1}
\usepackage{graphicx}

\begin{document}

\title{Similarities between Insect Swarms and Isothermal Globular Clusters}

\author{Dan Gorbonos}
\affiliation{Department of Chemical Physics, The Weizmann Institute of Science, P.O. Box 26, Rehovot, Israel 76100}
\author{Kasper van der Vaart}
\affiliation{Department of Civil and Environmental Engineering, Stanford University, Stanford, California 94305, USA}
\author{Michael Sinhuber}
\affiliation{Department of Civil and Environmental Engineering, Stanford University, Stanford, California 94305, USA}
\author{James G. Puckett}
\affiliation{Department of Physics, Gettysburg College, Gettsyburg, Pennsylvania 17325, USA}
\author{Nicholas T. Ouellette}
\affiliation{Department of Civil and Environmental Engineering, Stanford University, Stanford, California 94305, USA}
\author{Andrew M. Reynolds}
\affiliation{Rothamsted Research, Harpenden, AL5 2JQ, UK}
\author{Nir S. Gov}
\email{nir.gov@weizmann.ac.il}
\affiliation{Department of Chemical Physics, The Weizmann Institute of Science, P.O. Box 26, Rehovot, Israel 76100}

\begin{abstract}
Previous work has suggested that disordered swarms of flying insects can be well modeled as self-gravitating systems, as long as the ``gravitational'' interaction is adaptive. Motivated by this work we compare the predictions of the classic, mean-field King model for isothermal globular clusters to observations of insect swarms. Detailed numerical simulations of regular and adaptive gravity allow us to expose the features of the swarms' density profiles that are captured by the King model phenomenology, and those that are due to adaptivity and short-range repulsion. Our results provide further support for adaptive gravity as a model for swarms.
\end{abstract}

\maketitle

Insect swarms are a canonical example of collective animal behavior \cite{krause2002,sumpter2010}, displaying group-level cohesion and stability even in the presence of environmental noise \cite{okubo1986,kelley2013,attanasi2014,ni2015,sinhuber2019b,van2019mechanical}. But while in many other forms of collective animal motion, such as flocking, schooling, and herding \cite{sumpter2010}, the movement of individuals is coordinated, swarms are distinguished by their lack of globally aligned motion and the swarm state is not described by any order parameter.

It is thought that many swarming insect species such as {\em{Chironomus riparius}}, the midge species we consider here, interact predominantly via long-range acoustic sensing~\cite{aldersley2017emergent}. Indeed, a theoretical model that assumes that midges accelerate toward the sounds produced by other individuals has produced nontrivial results that are in good agreement with empirical observations \cite{gorbonos2016,gorbonos2017}. The acoustic field produced by flying insects has a monopole component whose intensity falls off according to an inverse-square law \cite{sueur2005}, a scaling that is similar to the way the gravitational pull between objects falls off with distance. Dipole and other higher-order multipole components decay more rapidly, and are thus weaker than the monopole component. It is therefore tempting to speculate, as Okubo \cite{okubo1986} and then Gorbonos \textit{et al.}~\cite{gorbonos2016} did, that midge swarms are analogous to $N$-body self-gravitating systems, where cohesion originates from the gravitational pull between the bodies compromising the system. A key contribution from ~\cite{gorbonos2016} was to incorporate the adaptive gain of typical biological sensors, which leads to results that are different from Newtonian gravitation but are in good agreement with empirical measurements of swarms.

The analogy between swarms and self-gravitating systems is appealing because it is well known that gravity can produce complex dynamical behavior from simple interactions---just as is thought to be the case for collective animal behavior \cite{vicsek1995}. Thus, by making a link between these two distinct systems, we can draw on the intuition built up from studying gravity to gain insight into collective animal behavior.

In this Letter, we examine this analogy more closely by studying the spatial variation of the number density and the velocity of individuals.
We compare to the classic King model for the mass distribution in isothermal globular clusters \cite{king1966} (see SI). Although we find some similarities with the observed swarms, there are also significant differences. These discrepancies reveal the limitations of a pure gravitational model for swarms. However, we find that the addition of adaptivity, which we can introduce only using numerical simulations, together with short-range repulsion can mitigate these limitations, producing mass distributions and dynamics that capture all the main features of the data. Thus, our results provide further support for adaptive gravity as a model for the swarm behavior.

\begin{figure}
\centering
\includegraphics[width=\linewidth]{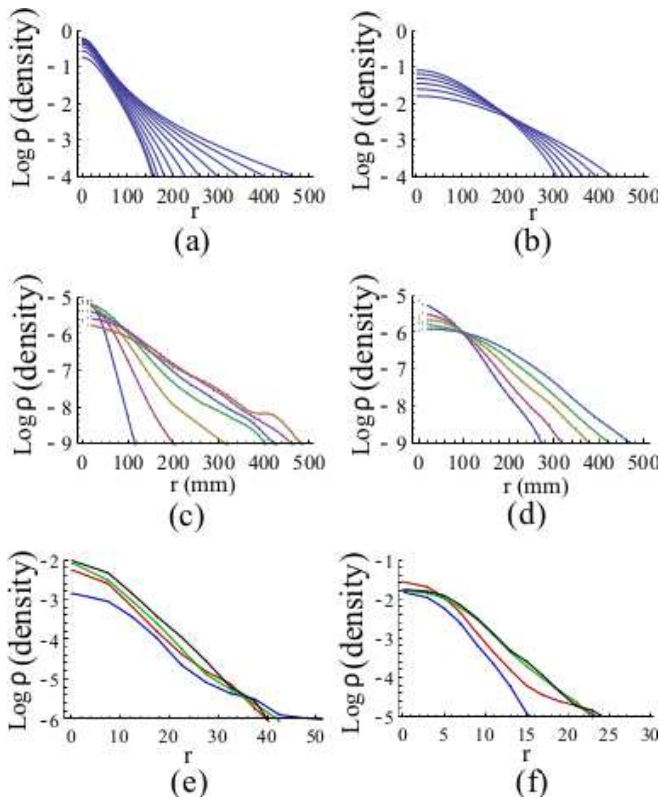}
\caption{\label{fig:density}
(a, b) Two families of density profiles computed from the King model (Eqs.~S15-S16) for different initial conditions, with shapes that are qualitatively similar to results from real swarms (c, d). (a) corresponds to the unstable branch, and (b) to the stable branch (see the discussion in the text).
(c, d) Two families of density profiles measured in real swarms in two different laboratory setups (see text).
(e) Density profiles of simulated swarms using the adaptive gravity model. Simulations were run using $N=12$ and $R_s=16.2$ (blue); $N=24$ and $R_s=11.8$ (red); $N=32$ and $R_s=11.5$ (green); and $N=48$ and $R_s=12$ (black).
(f) Density profiles of simulated swarms including adaptivity and short-range repulsion. Simulations were run using $N=12$ and $R_s=6$ (blue); $N=24$ and $R_s=6.7$ (red); $N=32$ and $R_s=8.2$ (green); and $N=48$ and $R_s=8.1$ (black).}
\end{figure}

We begin by considering the mass distributions predicted by the original King model. This model has been well studied previously. Chavanis \textit{et al.}~\cite{chavanis2015}, for example, described the solution space for globular clusters by fixing the cluster mass and varying the other system parameters, such as temperature, cluster radius, energy, and so forth. To adapt this model for swarms, we use a fixed inverse temperature $\beta$, which describes isothermal globular clusters. In addition to being simple to implement (see SI), this choice is motivated by the observation that swarms of different sizes empirically display roughly the same amount of kinetic energy per midge~\cite{puckett2014determining,reynolds2017}. For large globular clusters, a King model with fixed temperature gives a roughly constant kinetic energy per particle (independent of system size (Fig.~S1)). With this assumption, we compute the density profiles predicted by the King model (Fig.~\ref{fig:density}(a,b)), using Eqs.~S15 and S16.

\begin{figure}
\centering
\includegraphics[width=\linewidth]{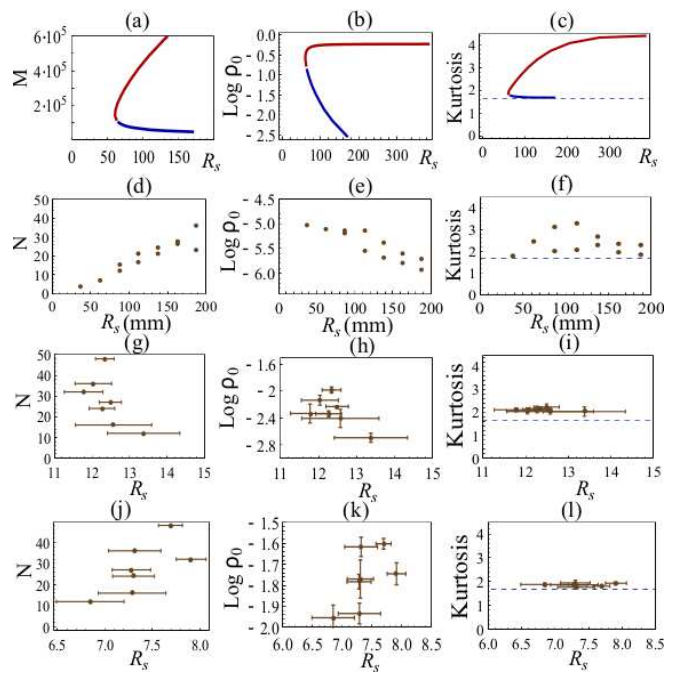}
\caption{\label{fig:panels}
(a-c) Total mass $M$ (a), density at the center $\rho_0$ (b), and kurtosis (c) as a function of the swarm size $R_s$ for a series of King model solutions. The blue line corresponds to the stable branch in the canonical ensemble and the red one to the unstable one. The kurtosis of a Gaussian distribution in three dimensions, $5/3$, is denoted by the horizontal dashed line. (d-f) The same quantities computed from the measured data for real swarms. (g-i) The same quantities computed for simulated swarms with adaptivity. (j-l) The same quantities computed for simulated swarms with adaptivity and short-range repulsion.}
\end{figure}

The King model predicts two distinct branches of solutions, which are clearly observed when we plot quantitative measures of the distributions such as the total mass $M$ (Fig.~\ref{fig:panels}(a), where we assume that all midges have the same mass), the density at the center $\rho_0$ (Fig.~\ref{fig:panels}(b)), and the kurtosis (Fig.~\ref{fig:panels}(c)) as functions of the overall swarm size $R_s$. Here, $R_s$ is defined as the mean distance of a midge from the center of mass of the swarm: $R_s\equiv\int_{0}^{\infty}r^3\rho(r)dr/\int_{0}^{\infty}r^2\rho(r)dr$, where $\rho(r)$ is the density profile. The two branches are termed ``stable'' and ``unstable'' (shown in blue and red, respectively, in Fig.~\ref{fig:panels}(a-c)), and are distinguished by the sign of the heat capacity (positive or negative, respectively) of the cluster in the canonical ensemble~\cite{chavanis2015}. It is found empirically that globular clusters reside on the unstable branch~\cite{chavanis2015}, presumably due to the strong destabilizing effects of ``slingshots''---that is, anomalously high acceleration events that arise due to close encounters.

In Fig.~\ref{fig:density}(c,d) we plot the empirical density profiles measured for laboratory midge swarms. Details of the laboratory setup and measurement protocols are given in refs.~\cite{kelley2013} and \cite{sinhuber2019}. The primary difference between the data in Fig.~\ref{fig:density}(c) and (d) is that the swarms in (c) were observed in a cubical laboratory enclosure measuring 91~cm on a side, while the enclosure for the swarms in (d) measured 122~cm on a side. Qualitatively, the density profiles from the swarms are similar in many aspects to those computed from the King model, though the agreement is not exact. To illuminate these similarities and differences further, we computed the same distribution measures for the swarms as for the King model, shown in Fig.~\ref{fig:panels}(d-f). The density in the center of the midge swarm $\rho_0$ slightly decreases with increasing swarm size, which qualitatively fits the stable branch of the King model (where this decrease is much steeper). The total number of midges (as a proxy for the total mass) increases slowly with the swarm radius, a feature exhibited by the unstable branch of the King model (the stable branch has the opposite relation). The kurtosis of the midge swarms falls in a range of values between the King model branches, and is significantly larger (i.e., the swarms have heavier tails than they would if they were Gaussian) than the density profiles along the King model stable branch (which are very close to Gaussian).

To attempt to reconcile the model predictions with the empirical results for real swarms, we consider two modifications to normal Newtonian gravity. First, since particles in the simulations have a strong tendency to develop anomalously high accelerations due to slingshots and thus to evaporate from the cluster, it is common to include a softening parameter $\epsilon$ to the gravitational force~\cite{brackbill1985}, so that
\begin{equation}
\vec{F}^{i}_{\epsilon,g} = C\sum_{j}\hat{r}_{ij}\frac{1}{|\vec{r}_i-\vec{r}_j|^2+\epsilon^2}, \label{epsilon}
\end{equation}
where $C$ is a constant with units of (force $\times$ length$^2$), $\vec{r}_i$ is the position vector for midge $i$, and $\hat{r}_{ij}$ is the unit vector pointing from midge $i$ to midge $j$. This ``epsilon''-gravity modifies the force at short distances, up to $|\vec{r}_i-\vec{r}_j|=O(\epsilon)$. 

In addition, we simulated our previously introduced adaptive-gravity model \cite{gorbonos2016}. In this case, the effective force felt by midge $i$ due to midge $j$ is given by
\begin{equation}
\vec{F}^{i}_{\mbox{\scriptsize eff}}=C\sum_{j}\frac{\hat{r}_{ij}}{|\vec{r}_i-\vec{r}_j|^2+\epsilon^2}\frac{R_{\mbox{\scriptsize ad}}^{-2}}{R_{\mbox{\scriptsize ad}}^{-2}+\sum_{k}\left(|\vec{r}_i-\vec{r}_k|^2+\epsilon^2\right)^{-1}} \label{feffmanySoft}
\end{equation}
The logic underlying this model is that the strength of the signal received by each midge should be renormalized by the total buzzing noise due to all the other midges, to mimic the typical adaptive gain of biosensors \cite{shoval2010}. $R_{\mbox{\scriptsize ad}}$ is the length scale over which this adaptivity occurs, and when $r_{ij} \gg \sqrt{N} R_{\mbox{\scriptsize ad}}$, such that the distance between a pair of midges is much larger than the adaptive range, the model reduces to epsilon-gravity (which itself reduces to Newtonian gravity when $\epsilon = 0$).

Finally, we also simulated the adaptive-gravity model with an additional short-range repulsion between the midges. This effect was empirically observed and measured in real swarms \cite{puckett2014searching}, and was implemented in an earlier simulation scheme \cite{gorbonos2016}. This type of effective interaction arises due to midges avoiding collisions while flying. Here we use a repulsive force with a cutoff to capture this effect, given by
\begin{equation}\label{frep}
    \vec{F}^{i}_{\mbox{\scriptsize rep}}=\left\{\,\begin{array}{cc}
     C\sum_{j}\left(\frac{1}{L_{c}^2}-\frac{1}{|\vec{r}_i-\vec{r}_j|^2}\right)\hat{r}_{ij}  &  |\vec{r}_i-\vec{r}_j|<L_c\\
     0    &  |\vec{r}_i-\vec{r}_j|\geq L_c
    \end{array}
    \right.
\end{equation}
where we take $L_c \approx 0.2\,R_{\mbox{\scriptsize ad}}$.

With these interaction laws, we considered several different swarm sizes $R_s$, as defined above. We used $R_{\mbox{\scriptsize ad}}=10 \approx R_s$, placing us in the regime where adaptivity affects the vast majority of the midges in the swarm. To explore the spatial distribution of mass in the adaptive gravity model we turned to numerical simulations. We used a scheme for performing $N$-body dynamics originally developed by Aarseth (see Appendix 4.B of \cite{binney2011}), which allows for accurate numerical integration using fourth-order equations of motion. A complete description of the numerical method is given in ref.~\cite{brackbill1985}. The acceleration of each mass in the simulation is computed by the direct summation of the forces due to the other $N-1$ bodies following the form of the force law. The scheme was designed to work efficiently for up to $N = 50$, well within the range of typical midge swarms \cite{kelley2013,sinhuber2019}. For further details about the simulation see the SI. Initially, particles were placed randomly in a simulation box of varying side lengths, in order to control the kinetic energy of the system. This was achieved (except for the epsilon-gravity system; see below) by varying the initial conditions of the simulations until the kinetic energy per particle was within $10\%$ of the desired value in the long-time limit (Fig.~S10). The initial velocities were zero, but under the influence of the forces the particles accelerated and eventually reached a quasi-stationary state (Figs.~S2-S4) that exhibits the distinct non-Gaussian velocity statistics (Fig.~S5) observed in real swarms \cite{kelley2013}. Note that we do not include explicit noise in the simulations, and the trajectories become ergodic due to the natural tendency of N-body gravitating systems to be chaotic.



We now compare the same quantitative features of the density profiles computed from the numerical simulations to those from the King model and the empirical swarm measurements. We start with the epsilon-gravity interaction (Eq.~(\ref{epsilon}), using $\epsilon=3.8$, which was much smaller than the swarm size $R_s$; Fig.~S12). Not surprisingly, some measures (Fig.~S12(b,c)) seem to agree well with the stable branch of the King model (Fig.~\ref{fig:panels}a-c), since close-encounter slingshots are greatly suppressed by the softened gravity. Other similarities include a decrease in the size $R_s$ and sharp decrease in $\rho_0$ for larger swarms. This force law therefore does not agree very well with the observed profiles of the midge swarms (Fig.~\ref{fig:panels}(d,e)), where the size increases with the number of particles and the density at the center shows a very weak decrease with size. Note that for epsilon-gravity, we were not able to keep the kinetic energy per particle constant when changing the overall number of particles in the swarm (Fig.~S11).

Figures \ref{fig:density}(e) and \ref{fig:panels}(g-i) show the results from adaptive-gravity simulations (Eq.~(\ref{feffmanySoft}), using $R_{ad}=10$). Note that the swarms cover a much smaller range of sizes ($R_s$), compared to the real (Figs.~\ref{fig:density}(c,d) and \ref{fig:panels}(d-f)) or epsilon-gravity swarms (Fig.~S12). This is due to the sharp increase in the density at the center when the number of particles increases combined with the fixed kinetic energy per particle, thereby maintaining a roughly constant $R_s$. We find that $\rho_0$ decreases with increasing $R_s$, similar to the observations in real swarms (as well as epsilon-gravity and the stable branch of the King model). The kurtosis is similar to the values seen in the real swarms. However, just as for epsilon-gravity and the stable branch of the King model, the biggest discrepancy with the real swarms is the relation between the number of particles (midges) and $R_s$, which we find in the simulations to be a decreasing function (Fig.~\ref{fig:panels}(g)) even though it is increasing in real swarms (Fig.~\ref{fig:panels}(d)).

We therefore also tested the addition of a short-range repulsion (Eq.~\ref{frep}) to the adaptive-gravity simulation (Figs.~\ref{fig:density}(f) and \ref{fig:panels}(j-l)). This additional ingredient makes $\rho_0$ roughly independent of  $R_s$ while not changing the kurtosis significantly. Both measures are in reasonable agreement with the real swarm data, considering the small range of swarm sizes in the simulations. The primary change as a result of the short-range repulsion is the appearance of an overall increasing relation between the number of particles and swarm size (Fig.~\ref{fig:panels}(j)). This feature, which is observed in the real swarms (Fig.~\ref{fig:panels}(d)), is absent from all the simulations that contain only long-range attractive interactions. This model therefore captures qualitatively all the main features of the density profiles of the midge swarms.

These results allow us to clearly delineate the properties of the midge swarm that are due to the long-range (adaptive) gravity interactions, namely the large kurtosis (heavy tails) and a slow decrease of the density at the swarm center for increasing swarm sizes. The short-range repulsion is responsible for inflating the size ($R_s$) of swarms with increasing number of midges, which is counter to the behavior of purely long-range interactions (both epsilon and adaptive gravity).

\begin{figure}
\centering
\includegraphics[width=\linewidth]{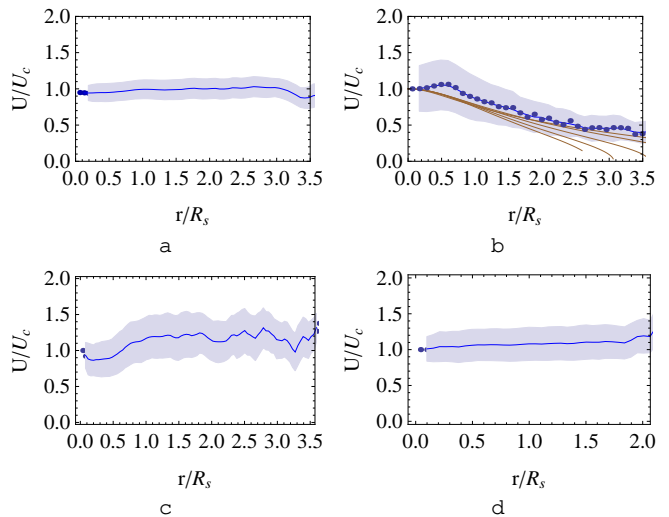}
\caption{\label{fig:speed_sims}
Mean speed as a function distance from the center of the swarm, normalized by the speed at the center. Data are shown for (a) real laboratory swarms, (b) epsilon-gravity, (c) adaptive gravity, and (d) adaptive gravity with short-range repulsion. The brown curves in (b) correspond to a family of King model solutions (Fig.~\ref{fig:density}a,b), computed numerically according to Eq.~($S23$) of the SI. In (d), we truncated the plot at $r\sim 2\,R_s$ since outside this radius the particles are not part of the cohesive region (i.e., the swarm) (see Fig.~(S9) of the SI).}
\end{figure}

So far we have considered the spatial density distribution of the swarm. We now turn to the spatial distribution of the velocities of the particles in the swarm, as another way to distinguish between the different models and compare to the observations from real swarms. We therefore computed the average speed of midges as a function of the distance $r$ away from the center of mass of the swarm using the dataset from the larger midge enclosure \cite{sinhuber2019}. As shown in Fig.~\ref{fig:speed_sims}(a), these speeds are essentially independent of position (as also noted in ref.~\cite{reynolds2017}). Thus, the mean kinetic energy of a midge is also statistically uniform in space (Fig.~S8(a)).

The simulated speed profile for epsilon-gravity is shown in Fig.~\ref{fig:speed_sims}b. We see that the velocity decreases rapidly with increasing radius from the swarm center, in good agreement with the profiles calculated from the King model but contrary to the data for real swarms. By contrast, the velocity profile computed using adaptive-gravity (Fig.~\ref{fig:speed_sims}(c)) is rather flat, with a small increase of speed for small $r$, presumably because the potential is too strongly softened in the swarm center. The velocity profile becomes even smoother when short-range repulsion is included (Fig.~\ref{fig:speed_sims}(d)), and is similar to the empirical results for real swarms. We find similar results for the standard deviation of the speed (Fig.~S8).

Finally, we considered the relation between mean acceleration and speed for swarms of different sizes (Fig.~S7)~\cite{reynolds2017}. We find that epsilon-gravity fails to reproduce the observed collapse of the curves for all swarm sizes, while the adaptive gravity captures this relation extremely well.

Taken together, our results provide further evidence that the interactions in insect swarms are well described by the adaptive-gravity framework, together with a short-range repulsion. This highlights that two key ingredients---long-range interactions and adaptivity---are essential. This model naturally captures many of the unusual properties of swarms, some of which have surprising similarities to globular clusters. Adaptivity, however, also induces additional effects. When the binding to the swarm is adaptive, the forces on one midge due to the other midges are not additive, and as a consequence the sum of the forces felt by all of the midge need not vanish, as it must for Newtonian gravity~\cite{gorbonos2016}. The center of mass of the swarm can therefore experience accelerations \cite{reynolds2016}. Such fluctuations have the potential to change fundamentally the characteristics of individual flight patterns; for example, Reynolds and Ouellette \cite{reynolds2016} showed that center of mass fluctuations allow for the emergence of L{\' e}vy flight patterns. Centre-of-mass movements may also help to stabilize insect swarms against environmental perturbations \cite{reynolds2019origin}.

These results suggest that it may be fruitful in the future to consider modifications of the interactions within the King model as a general framework for describing the behavior of active systems with long-range interactions, whatever their form. The proper analytical treatment of a King model augmented with an adaptive-gravity interaction law remains as a challenge for future studies. Finally, future work on swarm modeling in particular should explore the effects of adding self-propulsion and stochasticity (noise), both of which are absent from the current models but certainly present for real insects.

\begin{acknowledgments}
We thank Tsvi Piran, Sverre Aarseth, and Tal Alexander for useful discussions. The research at Stanford was sponsored by the Army Research Laboratory and accomplished under grant no.~W911NF-16-1-0185. The views and conclusions in this document are those of the authors and should not be interpreted as representing the official policies, either expressed or implied, of the Army Research Laboratory or the U.S. government. K.v.V. acknowledges support from an Early Postdoc Mobility fellowship from the Swiss National Science Foundation, and M.S. acknowledges support from the Deutsche Forschungsgemeinschaft under grant no.~396632606. The work at Rothamsted forms part of the Smart Crop Protection (SCP) strategic programme (BBS/OS/CP/000001) funded through the Biotechnology and Biological Sciences Research Council's Industrial Strategy Challenge Fund. NSG is the incumbent of the Lee and William Abramowitz
Professorial Chair of Biophysics.
\end{acknowledgments}

\bibliography{density}

\end{document}


\bibliographystyle{plainnat}

\title{Supplementary Information: ``Similarities between Insect Swarms and Isothermal Globular Clusters''}
\author{Dan Gorbonos$^1$, Kasper van der Vaart$^2$, Michael Sinhuber$^2$, James G. Puckett$^3$, Andrew M. Reynolds$^4$, Nicholas T. Ouellette$^2$ and Nir S. Gov$^1$\footnote{Correspondence and requests for materials should be addressed to N.S.G. (email: nir.gov@weizmann.ac.il).}}
\affiliation{$^1$Department of Chemical Physics, The Weizmann Institute of Science, P.O. Box 26, Rehovot, Israel 76100 \\
$^2$Department of Civil and Environmental Engineering, Stanford University, Stanford, California 94305, USA \\
$^3$Department of Physics, Gettysburg College, Gettsyburg, Pennsylvania 17325, USA\\
$^4$ Rothamsted Research, Harpenden, AL5 2JQ, UK}

\maketitle

\setcounter{equation}{0}  
\setcounter{figure}{0}

\subsection{The King Model}

 Most of the globular clusters are well-fitted by the King model~\cite{King}, since
 it captures the basic structure and main observed features of them~\cite{Chavanis:2014woa}. This model is one of a series of models that assume Maxwell–-Boltzmann distribution of the velocities, although it is well-known that a self-gravitating system cannot be in thermodynamical equilibrium. This fact is a result of the long-range nature of the interactions, and the Boltzmann entropy which has no maximal value in an unbounded domain. Therefore self-gravitating systems, as part of their out-of-equilibrium state, have a strong tendency to evaporate over time. Since evaporation is a slow process, the system can be considered to be in a quasi-stationary state for times that are much shorter than the evaporation time, and the Maxwell-–Boltzmann distribution can therefore serve as a good approximation.

 In order to write a quasi-stationary expression for the distribution $f(\vec{r},\vec{v})$ (a function of the position and the velocity that does not depend on time) we start from the colisionless Boltzmann equation:
  \be \label{Boltz}
  \vec{v}\cdot\nabla f=\nabla V\cdot\frac{\partial f}{\partial \vec{v}},
  \ee
  where $V=V(r)$ is a spherically symmetric gravitational potential.
  The density is defined by
  \be
  \rho=\int f d^3\vec{v}, \label{den}
  \ee
  and it determines the gravitational potential through the Poisson equation
  \be
 \triangle V = 4\,\pi\,C\,\rho,\label{Poisson}
  \ee
  where $C$ is a constant with dimensions of $mass \cdot length^{3}/time^{2}$ which for gravitational force is the gravitational constant.
  A simple solution to this set of equations ((\ref{Boltz}) and (\ref{Poisson})) is given by the Maxwell-–Boltzmann distribution with a singular power-law behavior in the spatial part:
  \be \label{singular}
   f(\vec{r},\vec{v})\sim\,\frac{e^{-\frac{\beta\,\vec{v}^2}{2}}}{r^2},
 \ee
  where $\beta$ is a positive constant which is identified with the inverse temperature.

  This solution is singular at the origin, unbounded and has an infinite mass since
  \be
  M(r)\propto r.
 \ee
  The singularity at the origin can be fixed by introducing regular boundary conditions:
  \be \label{boundary1}
  V(0)=const, \quad V'(0)=0,
 \ee
 but it does not fix the asymptotic behavior.
  Numerical solutions of equations ((\ref{Boltz}) and (\ref{Poisson})) with boundary conditions (\ref{boundary1}) are solutions of isothermal spheres (since $\beta$ is constant). Far from the center $r\gg (C\,\rho(0)\,\beta)^{-\frac{1}{2}}$ the isothermal sphere solutions approach the singular solution (\ref{singular}). Although the solution is not singular at the center, the solution is still unbounded and has an infinite mass (see for instance~\cite{padmanabhan2000theoretical}, p. 480-484).

 The steady-state mass distribution that solves the Poisson-Boltzmann equations is singular, proportional to $\rho(r)\propto 1/r^2$. To find physical solutions, it is necessary to introduce a cut-off. The assumption in the King model is that a cluster cannot keep stars whose velocity exceeds a finite escape velocity. Therefore the velocity distribution should drop to zero at a finite limiting velocity. As a result, the cluster has a finite radius. Different cut-offs in the velocity distribution give different limiting radii. Observationally the finite boundary is set by the tidal forces of the galaxy with which the globular cluster is associated, that become dominant at a certain distance from the cluster center. The velocity distributions are simply the Maxwell–-Boltzmann distributions minus a constant. Let us assume the following form at the origin:
 \be \label{centervel}
 f(0,\vec{v})=
 \left\{\!\!
   \begin{array}{cc}
    \alpha\,\(e^{-\frac{\beta v^{2}}{2}}-e^{-\frac{\beta v_{e}(0)^{2}}{2}}\) & v\leq v_{e} \\
     0 & v > v_{e}
   \end{array}
 \right.
 \ee
 where $\alpha$ is a positive constant.
 This way the fastest particles $v > v_{e}(0)$ are subtracted from the distribution.
 The distribution of the velocities at $r=0$ is defined by two parameters: $\beta$ (determines the width of the distribution) and $v_e(0)$, the escape velocity.


  According to Jeans' theorem any steady state solution of the collisionless Boltzmann equation is a function of the integrals of motion (see for instance~\cite{binney2011galactic}, p. 220). Since energy is the integral of motion in a star cluster the distribution function can be written as a function of the energy $E$:
 \be
 f(\vec{r},\vec{v})=f(E)
 \ee
 We express the cut-off as an escape energy $E_{e}$ such that $f=0$ for $E \geq E_{e}$.
 The escape energy corresponds to an escape velocity at a point $r$ through:
 \be \label{escape}
 E_{e}=\frac{v_{e}(r)^2}{2}+V(r).
 \ee

 \be
 \label{KingDist}
 f(E)=\begin{cases}
     \alpha\,e^{\beta\(V(0)-E_{e}\)}\left[e^{-\beta(E-E_{e})}-1\right], & E<E_{e}\\
     \\
     0 & E\geq E_{e}.
   \end{cases}
 \ee
 where $\alpha$ is a positive constant.

 Substituting the distribution function into $(\ref{den})$ and changing variables give us:
 \bea
 w &\equiv& \(\frac{\beta}{2}\)^{\frac{1}{2}}v \label{change1}\\
 \chi (r) &\equiv& \beta \(E_{e}-V(r)\) \label{change2}
 \eea
 we get
 \be
 \rho=\begin{cases} \label{density}
     4\,\pi\,\alpha\,e^{-\chi(0)}\,(\frac{2}{\beta})^{\frac{3}{2}}I(\chi), & r<R_{e} \\
     \\
     0 & r\geq R_{e}.
     \end{cases}
 \ee
 where
 \be
 I(x)\equiv \int_{0}^{\sqrt{x}}\(e^{x-w^2}-1\)w^2\,dw,\ee
 and $R_{e}$ is the radius of the cluster such that $v_e(R_e)=0$, and $\chi(R_e)=0$. In other words, any star that succeeds to reach $R_{e}$, escapes the cluster and it is not a part of the cluster any more.

 When we substitute the density (\ref{density}) in the Poisson equation (\ref{Poisson})
 we get the fundamental equation of the King model
 \be
 \frac{1}{\zeta^2}\frac{d}{d\zeta}\(\zeta^2\frac{d\chi}{d\zeta}\)=-\frac{I(\chi)}{I(k)} \label{fund}
 \ee
 with the boundary conditions
 \be
 \chi(0)=k,\quad \chi'(0)=0, \label{initialCon}
 \ee
 where the rescaled dimensionless distance is
 \be
 \zeta \equiv r (4\,\pi\,C\,\beta \,\rho(0))^{\frac{1}{2}}.
 \ee
 The set of King solutions for the density is a one parameter family of solutions which are conveniently parameterized by
 $k$,which is the rescaled potential at the center.

 Using the expression for the density~(\ref{density}) and the fundamental equation~(\ref{fund}), we can write the mass in the following way:
 \be
 M(r)=4\,\pi\int_0^{r}\rho(r')\,r'^2\,dr'=-\(4\,\pi\,\rho(0)\)^{-\frac{1}{2}}\,(\beta\,C)^{-\frac{3}{2}}\,\zeta^2\chi'(\zeta).
 \ee
 Note that here too, in the limit $k\rightarrow\infty$, the singular solution $(\ref{singular})$ is recovered (see~\cite{Chavanis:2014woa}).

\begin{figure}[htb]
\centering
\includegraphics[width=0.6\linewidth]{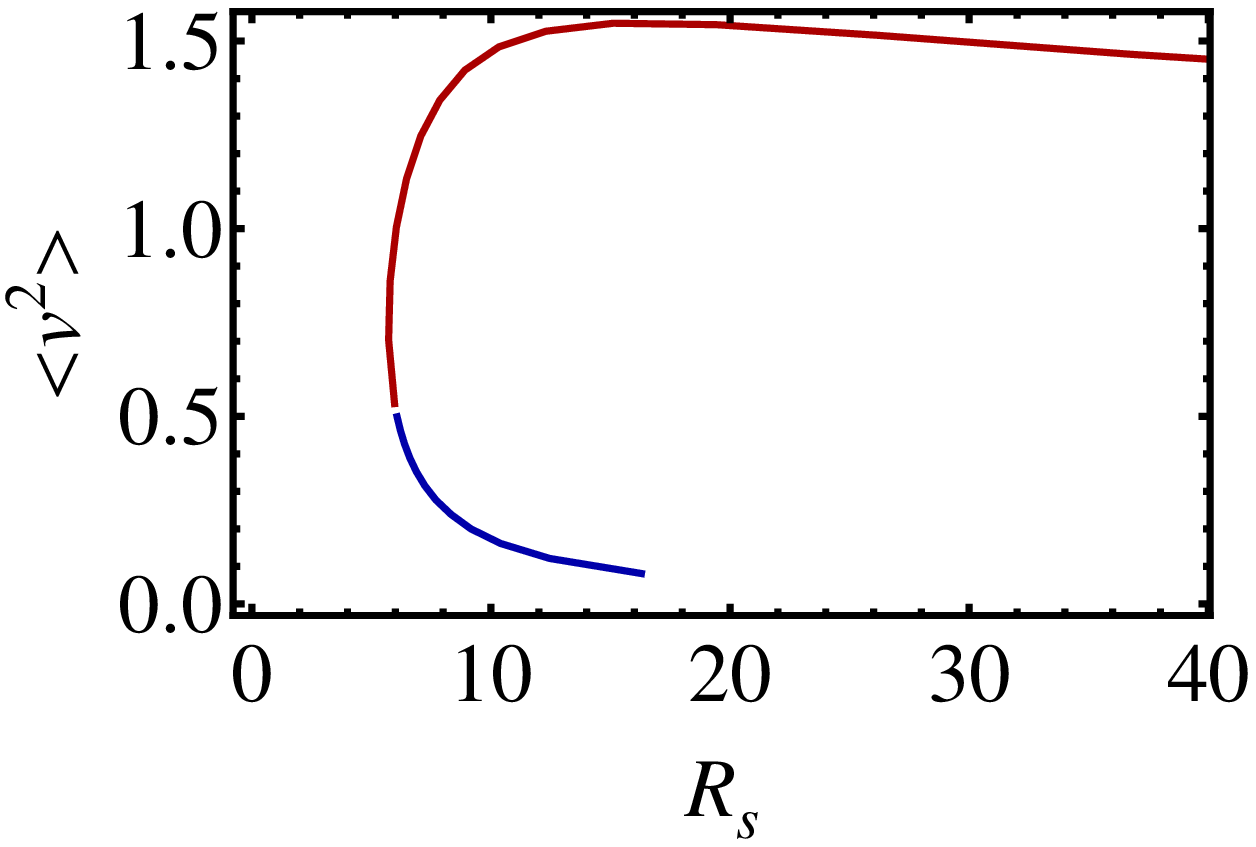}
\caption{  The average kinetic energy (per particle) as a function of $R_s$ for a series of King solutions. The blue and the red lines stand for the stable and unstable branches in the canonical ensemble respectively. Note that in the limit of large clusters the average kinetic energy is approximately independent of the cluster size.}\label{AverageKinetic}
\end{figure}

 In Fig.~\ref{AverageKinetic} we give the average kinetic energy (per unit mass) as a function of $R_s$ for various King solutions. The average kinetic energy (per unit mass) $<E_{k}>$ is given by:
\be
<E_{k}>=\frac{1}{2}<v^2>=\frac{1}{2}\int\!\!\int v^2\,f(\vec{r},\vec{v})d^{3}\vec{r}\,d^{3}\vec{v},
\ee
when $f(\vec{r},\vec{v})$ is normalized so that
\be
\int\!\!\int v^2\,f(\vec{r},\vec{v})d^{3}\vec{r}\,d^{3}\vec{v}=1.
\ee
In the case of an isotropic distribution of velocities and spherical symmetry we get:
\be \label{firstv2}
v^2(r)=\int_{0}^{v_e(r)} v^4\,f(r,v)\,dv\left/\int_{0}^{v_e(r)} v^2\,f(r,v)\,dv\right..
\ee
Substituting ($\ref{KingDist}$) and changing variables according to $(\ref{change1})$ and $(\ref{change2})$ we obtain:
\be
v^2(r)=\frac{2}{\beta}\int_{0}^{\sqrt{\chi}} \left(e^{\chi-w^2}-1\right)w^4\,dw\left/I(\chi)\right.
\ee
for $r<R_{e}$.
Integration by parts gives us:
\be \label{lastv2}
v^2(r)=\frac{6}{5\beta}\frac{\int^{\sqrt{\chi}}_{0}w^6\,e^{-w^2}dw}{\int^{\sqrt{\chi}}_{0}w^4\,e^{-w^2}dw}.
\ee
Repeating the steps in Eqs. ($\ref{firstv2}$)-($\ref{lastv2}$) with additional integration over the spatial directions gives the average kinetic energy:
\be
<E_{k}>=\frac{1}{2}<v^2>=\frac{3}{5\beta}\frac{\int_{0}^{R_{e}}r^2dr\,e^{\chi(r)}\int^{\sqrt{\chi}}_{0}w^6\,e^{-w^2}dw}{\int_{0}^{R_{e}}r^2dr\,e^{\chi(r)}\int^{\sqrt{\chi}}_{0}w^4\,e^{-w^2}dw}.
\ee
In Fig.~(\ref{AverageKinetic}) $<v^2>$ is given for a series of solutions of Eq.~(\ref{fund}) with different values of $k$. For simplicity we take $\beta=3/5$ in this figure.

The local velocity dispersion of a spherically symmetric distribution function is defined by~\cite{Chavanis:2014woa}
\be
\sigma^2(r)=\frac{v^2(r)}{3}.
\ee
The average velocity is given by
\be
<v>=\frac{\int_{0}^{R_e}\,r^2\,dr\int_{0}^{v_e(r)}v^3\,f(r,v)dv}{\int_{0}^{R_e}\,r^2\,dr\int_{0}^{v_e(r)}v^2\,f(r,v)dv},
\ee
and then we can write the dispersion normalized by the average velocity:
\be \label{sigmanormalizedKing}
\frac{\sigma(r)}{<v>}=\frac{4}{3\sqrt{5}}\sqrt{\frac{\int^{\sqrt{\chi}}_{0}w^6\,e^{-w^2}dw}{\int^{\sqrt{\chi}}_{0}w^4\,e^{-w^2}dw}}\left/\frac{\int_{0}^{R_e}r^2dr\,e^{\chi(r)}\int_{0}^{\sqrt{\chi}}w^5e^{-w^2}dw}{\int_{0}^{R_e}r^2dr\,e^{\chi(r)}\int_{0}^{\sqrt{\chi}}w^4e^{-w^2}dw}\right.
\ee

 \newpage

\subsection{The Time Evolution of the Swarm in the Simulation}
\begin{figure}[htb]
\centering
\includegraphics[width=0.45\linewidth]{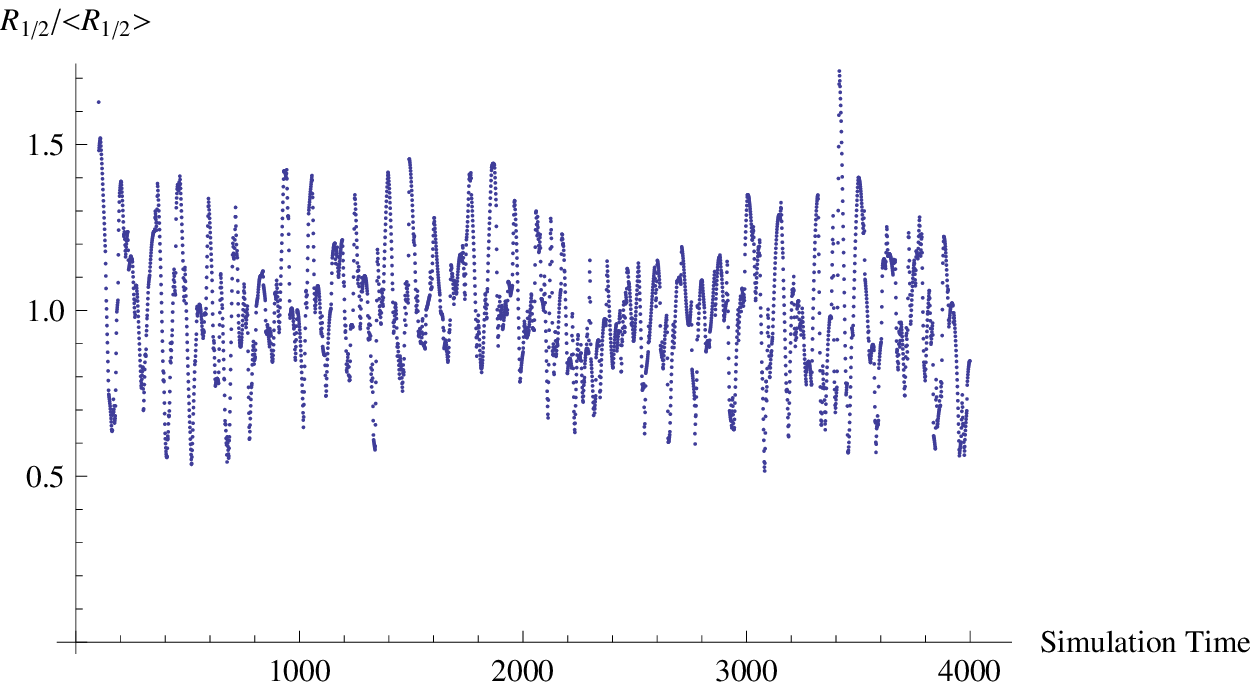}
\caption{\label{halfRadius}  The half-mass radius (normalized by its mean value) as a function of simulation time for a swarm with adaptivity ($R_{ad}=5$) and 50 individuals. The half-mass radius is defined as the minimal radius that bounds half of the individuals in the swarm. Over the sampled time period the swarm seem to be stable for the randomly chosen initial conditions, which means that the evaporation takes place on a longer time scale.}
\end{figure}

\begin{figure}[htb]
\centering
\includegraphics[width=0.45\linewidth]{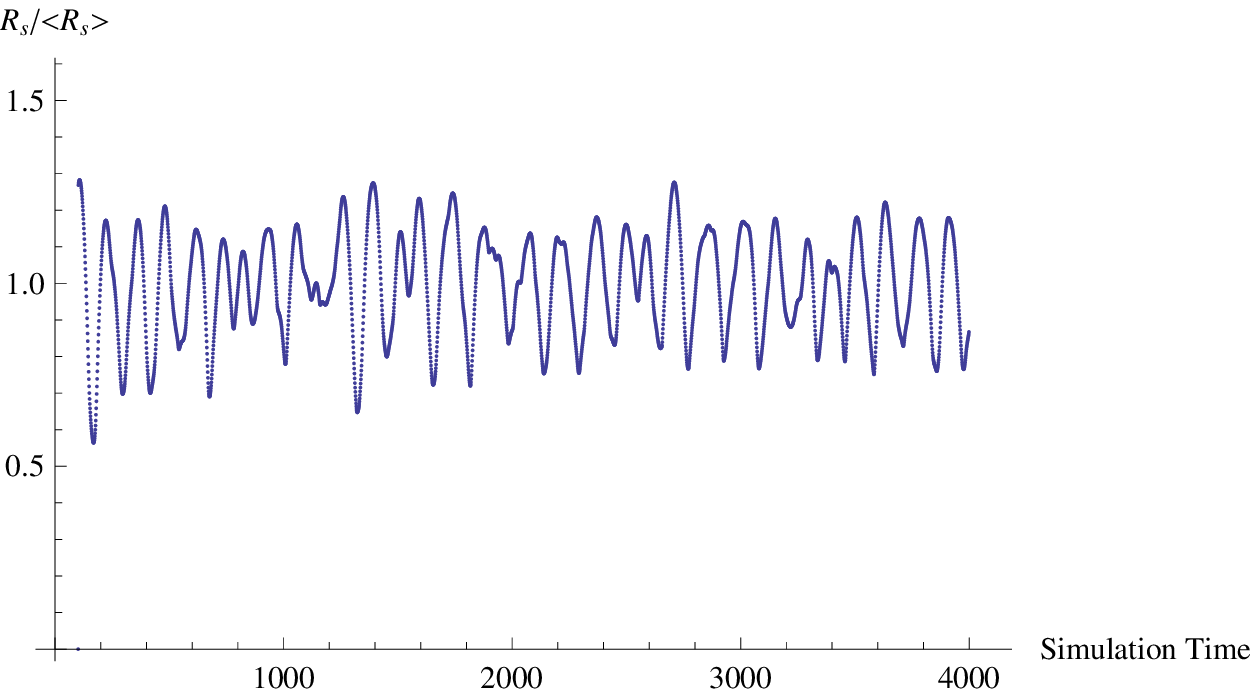}
\caption{  \label{SizeTime} The size of the swarm (normalized by its mean value) as a function of simulation time for the same swarm as in Fig.~\ref{halfRadius} ,where the size is defined as the mean distance from the center of the mass of the swarm.}
\end{figure}

\begin{figure}[hbt]
\centering
\includegraphics[width=0.45\linewidth]{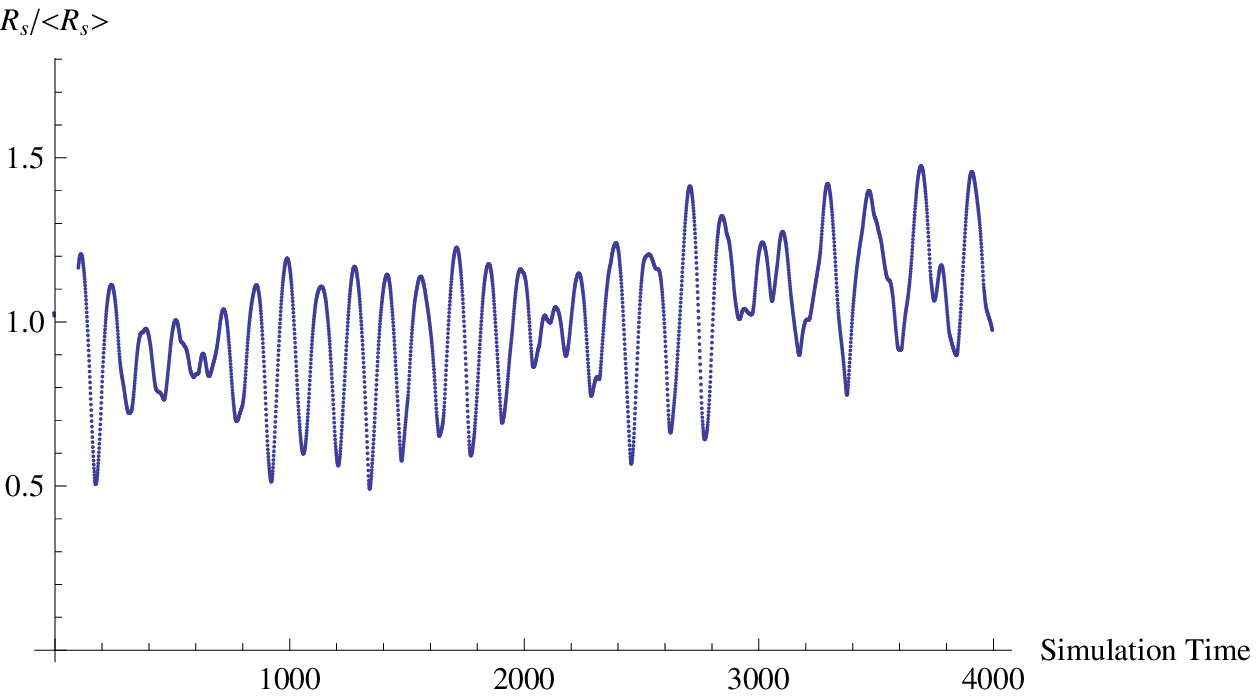}
\caption{\label{Evapor} The size of the swarm (normalized by its mean value) as a function of simulation time for a swarm with adaptivity ($R_{ad}=5$) and 50 individuals. Here, in this example, different initial conditions create an evaporation at earlier time than the previous example in Fig.~\ref{SizeTime}. In the examples that we considered in the paper we did not include time segments similar to this one, that showed a significant evaporation.}
\end{figure}

Here we show a typical oscillatory behavior as a function of time of the half-mass radius (Fig.$~\ref{halfRadius}$) and the size of the swarm (mean distance from the center of the swarm) Fig.~$\ref{SizeTime}$ for initial conditions that do not cause evaporation (over the time segment that is presented here). In Fig.~$\ref{Evapor}$ we show that the same swarm with different initial conditions evaporates over the same time segment (Fig.~$\ref{Evapor}$). Such cases were not considered as stable in this paper, and were ignored.

\newpage

\subsection{The velocity and acceleration distributions}
\begin{figure}[h]
\centering
\includegraphics[width=0.5\linewidth]{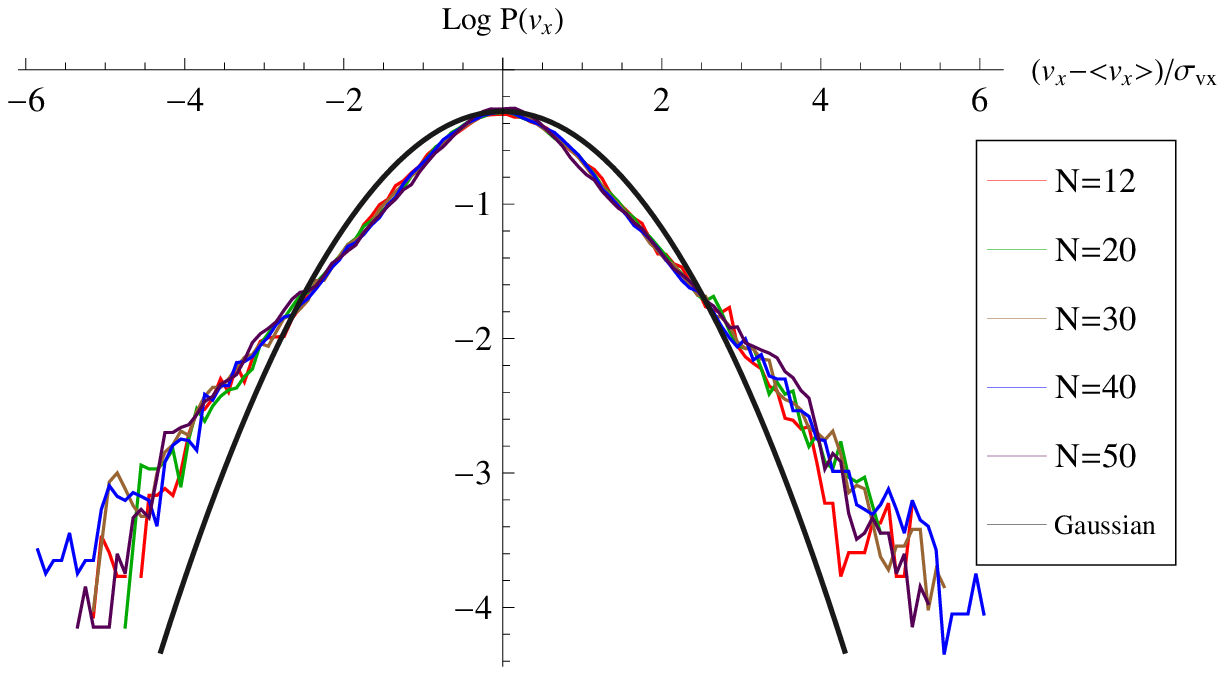}
\caption{\label{velxDist} The x-component velocity distribution normalized by the standard deviation for different simulated swarms with adaptivity ($R_{ad}=5$). A reference Gaussian curve is shown in black.}
\end{figure}

\begin{figure}[h]
\centering
\includegraphics[width=0.6\linewidth]{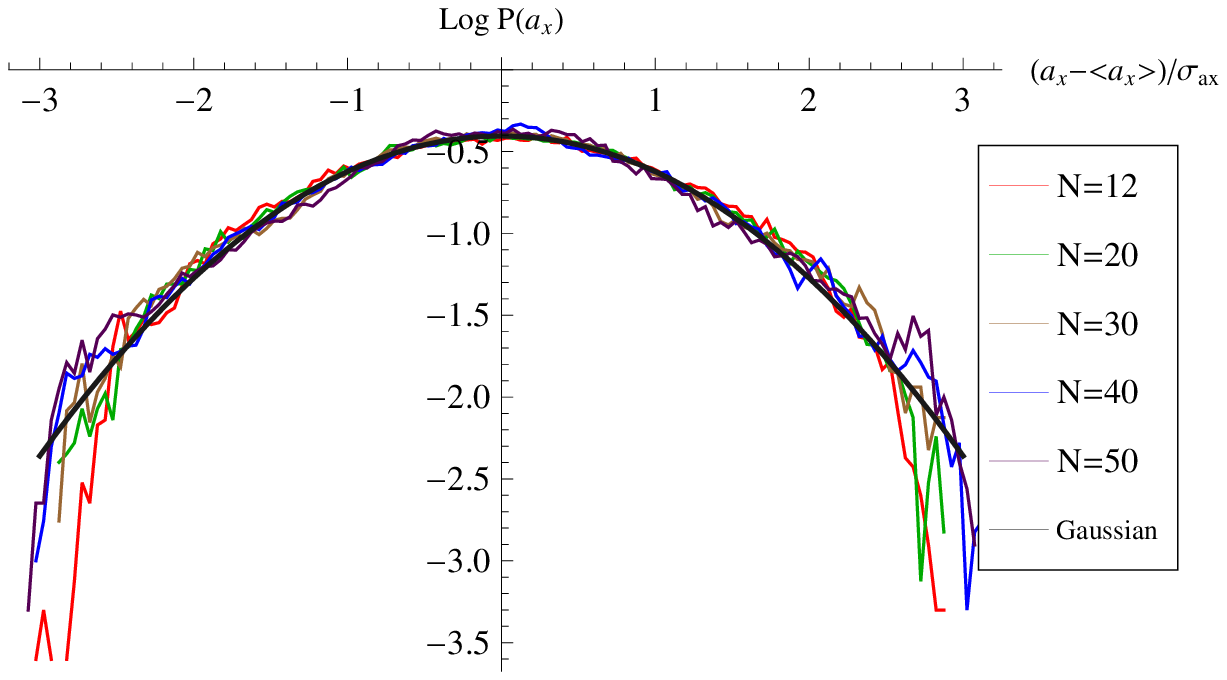}
\caption{\label{AccxDist} The x-component acceleration distribution normalized by the standard deviation for different simulated swarms with adaptivity ($R_{ad}=5$). The acceleration distribution is bounded and cannot reach values which are too far from the average value. This is a result of the addition of the adaptivity and the softening parameter $\epsilon$ to the simulation, which is needed in order to slow down the evaporation of the swarm. A reference Gaussian curve is shown in black.}
\end{figure}

The distribution of a single component of the velocity in the simulation is given in Fig.~$\ref{velxDist}$ for various swarm sizes. In Fig.~$\ref{AccxDist}$ we give the similar distribution of a single component of the acceleration. The structure of the tails is similar to the same distributions that were obtained from the laboratory data~\cite{kelley2013}.


\subsection{The relation between mean acceleration and speed}

\begin{figure}[h]
\centering
\includegraphics[width=0.5\linewidth]{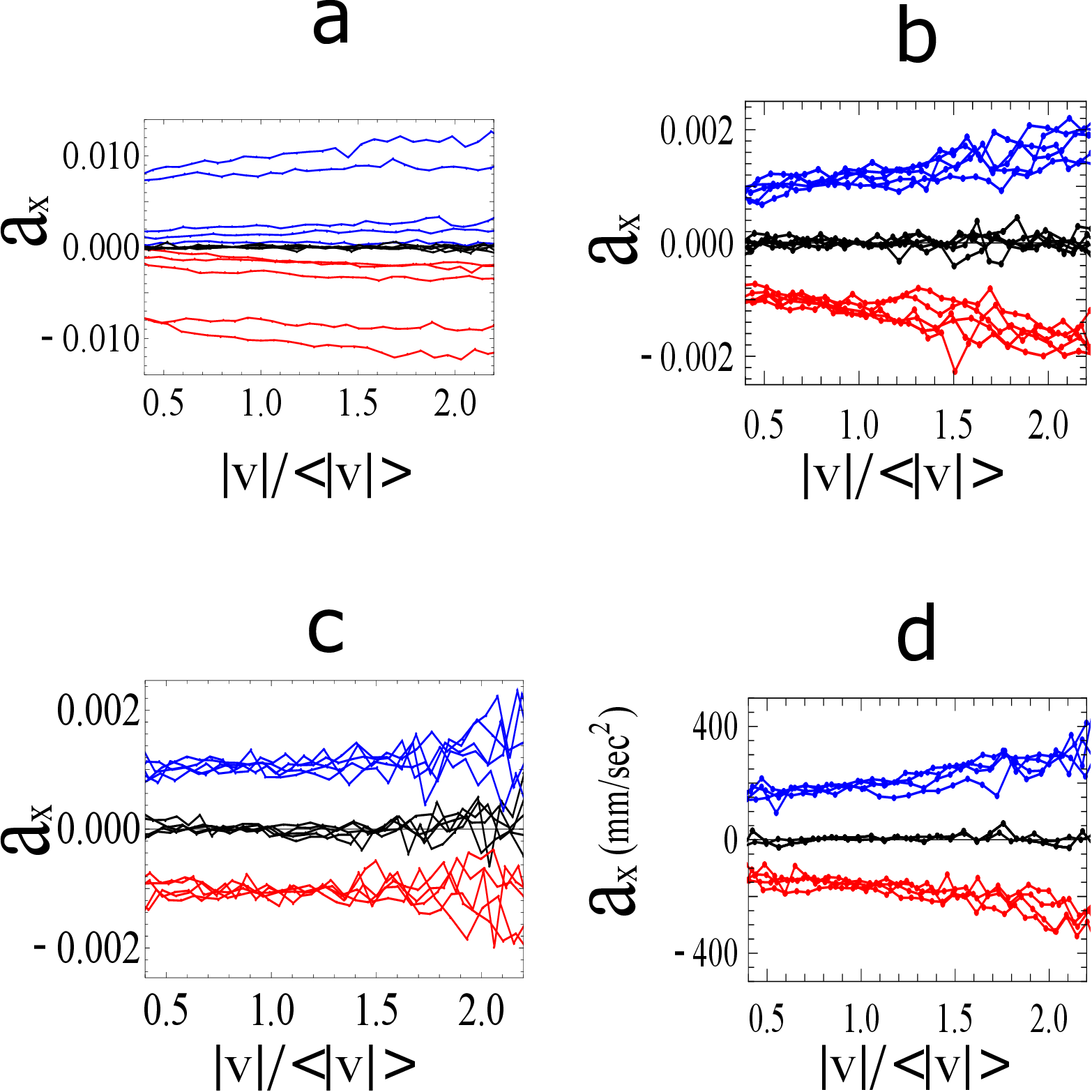}
\caption{\label{avsv} The mean value of a single component of the acceleration vs. the speed (normalized by the average speed) of the left hemisphere of different swarm (blue), right hemisphere (red) and the black
 lines show accelerations for both hemispheres, which (as required by symmetry) is close to zero. (a) - $\epsilon$-gravity (The two higher accelerations corves actually correspond to cases with lower $R_s$, from the upper to the lower we have $R_s=16.23,3.16,19.53,18.37,30.10$) (b) -  adaptivity (c) adaptivity with repulsion (d) experimental data. }
\end{figure}

It was observed in~\cite{Reynolds2017} that if we take the mean acceleration of the right or left hemisphere of the swarm separately, it's absolute value is a monotonously increasing function of the speed. The experimental plot from~\cite{Reynolds2017} for various sizes of swarms is reproduced here in Fig.~\ref{avsv}d where the speed is normalized with respect to its mean value. Comparing with the simulations results (for $\epsilon$-gravity in Fig.~\ref{avsv}a, adaptivity in Fig.~\ref{avsv}b and adaptivity with short-range repulsion in Fig.~\ref{avsv}c) for swarms of different sizes, we see that in the cases with adaptivity (with and without repulsion) all the graphs collapse at low velocities to have approximately the same acceleration, when in the $\epsilon$-gravity case there are various values of acceleration at low velocities, and it does not agree with the experimental data in Fig.~\ref{avsv}d. One can also see that with repulsion there is more noise at high velocities. The fitting that we see here between adaptivity and the experimental data suggests that adaptive-gravity naturally explains the dependence of the acceleration on the mean velocity, without the need to invoke an additional explicit velocity-dependent forces~\cite{Reynolds2017}.



\subsection{The standard deviation of the speed in the simulations and the experiments}

\begin{figure}[h]
\centering
\includegraphics[width=\linewidth]{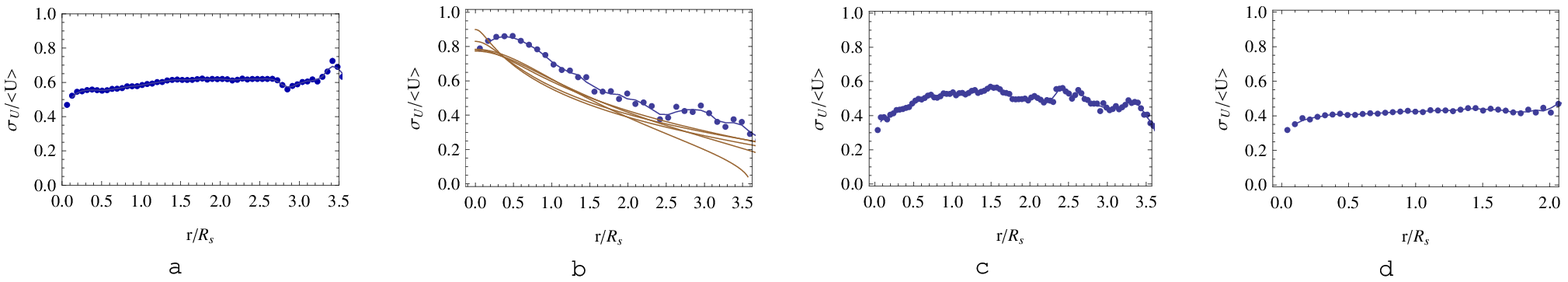}
\caption{\label{FlatSpeed2} The standard deviation of the speed normalized by the average speed in the swarm vs. the distance from the center of mass normalized by the swarm size $R_{s}$. (a) - Experimental results (b) - Simulation results for $\epsilon$-gravity (without adaptivity). The brown curves correspond to a family of King solutions computed numerically according to Eq.~($\ref{sigmanormalizedKing}$). (c)- Simulation results with adaptivity (d) - Simulation results with adaptivity and repulsion.
}
\end{figure}

\begin{figure}[h]
\centering
\includegraphics[width=\linewidth]{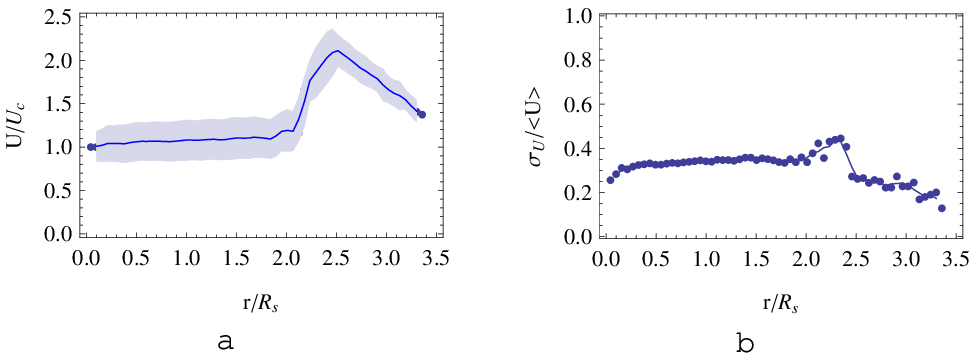}
\caption{\label{FlatSpeed3} Simulations with adaptivity and repulsion - Results till $r \sim 3.5\,R_s$. The part $2\,R_s<r<3.5\,R_s$ was omitted in Fig.~3d and Fig.~\ref{FlatSpeed2}d since it describes particles that are not cohesive as a result of the repulsion. Here we show two of the characters (a) The mean speed vs. the distance from the center of the swarm, normalized by the speed at the center.
(b) The standard deviation of the speed normalized by the average speed in the swarm vs. the distance from the center of mass normalized by the swarm size $R_{s}$.
We see that for $r>2\,R_{s}$ the velocities become up to twice higher, the standard deviation of the velocities becomes twice lower and the density is less than $10^{-5}$ of the density at the center (Fig.~1). Therefore this region probably describes particles that were repelled out of the swarm at high velocities (higher than the escape velocity).
}
\end{figure}

\subsection{Kinetic energy in the simulation}

The kinetic energy of each midge in the laboratory is approximately constant and does not depend on the density or the size of the swarm. In the simulation, since we are interested only in the mutual forces, we do not require constant speed for the particles, and therefore the kinetic energy per individual is different for various numbers of particles and initial conditions. We chose initial conditions that give us approximately the same kinetic energy per individual in the case of simulations with adaptivity and adaptivity with repulsion (up to 10\%). Without adaptivity we could not change significantly the kinetic energy by varying the initial conditions and it was determined mainly by the number of particles in the simulation.

\begin{figure}[h]
\centering

\includegraphics[width=0.6\linewidth]{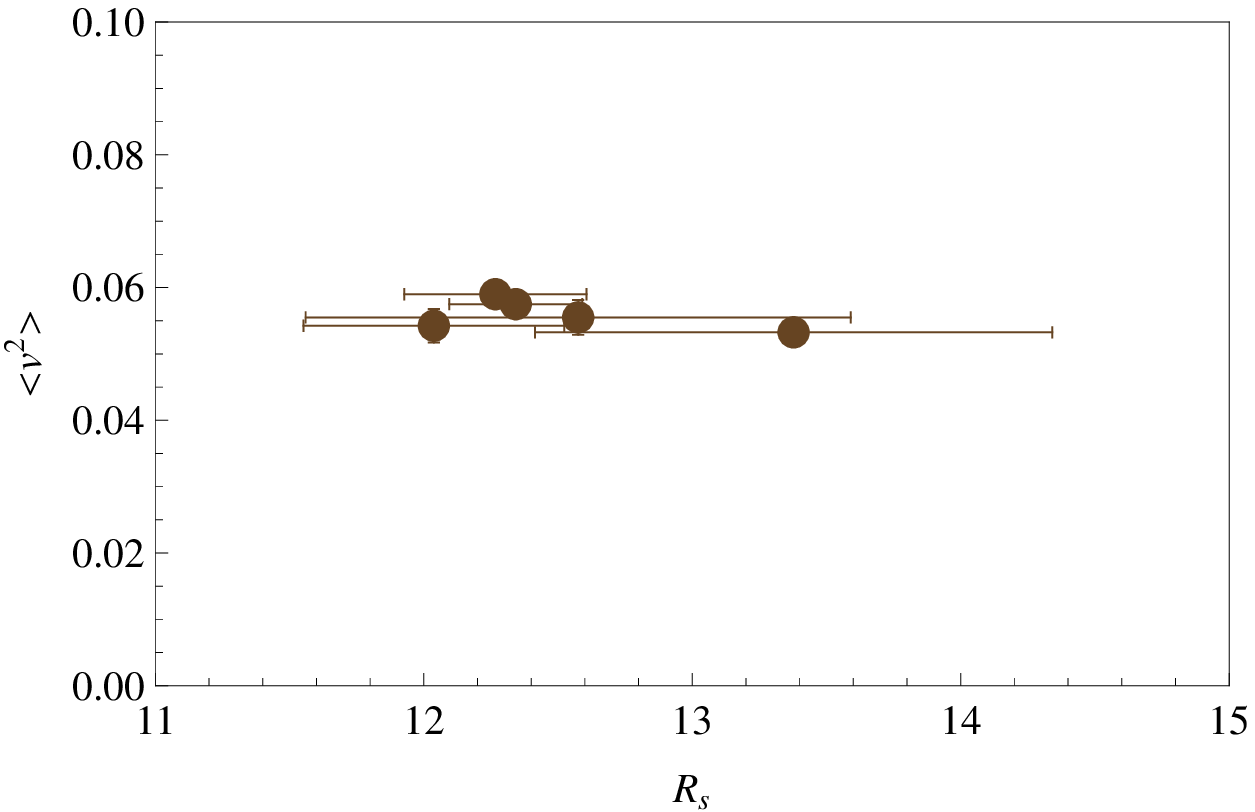}
\caption{\label{KEvsRs} The kinetic energy as a function of the size of the swarm $R_s$ for cases with adaptivity (Fig.~2).
Here the size of the swarm does not change much but the number of individuals change from 12 to 48 (Fig.~2).
}
\end{figure}

\begin{figure}[h]

\centering
\includegraphics[width=0.6\linewidth]{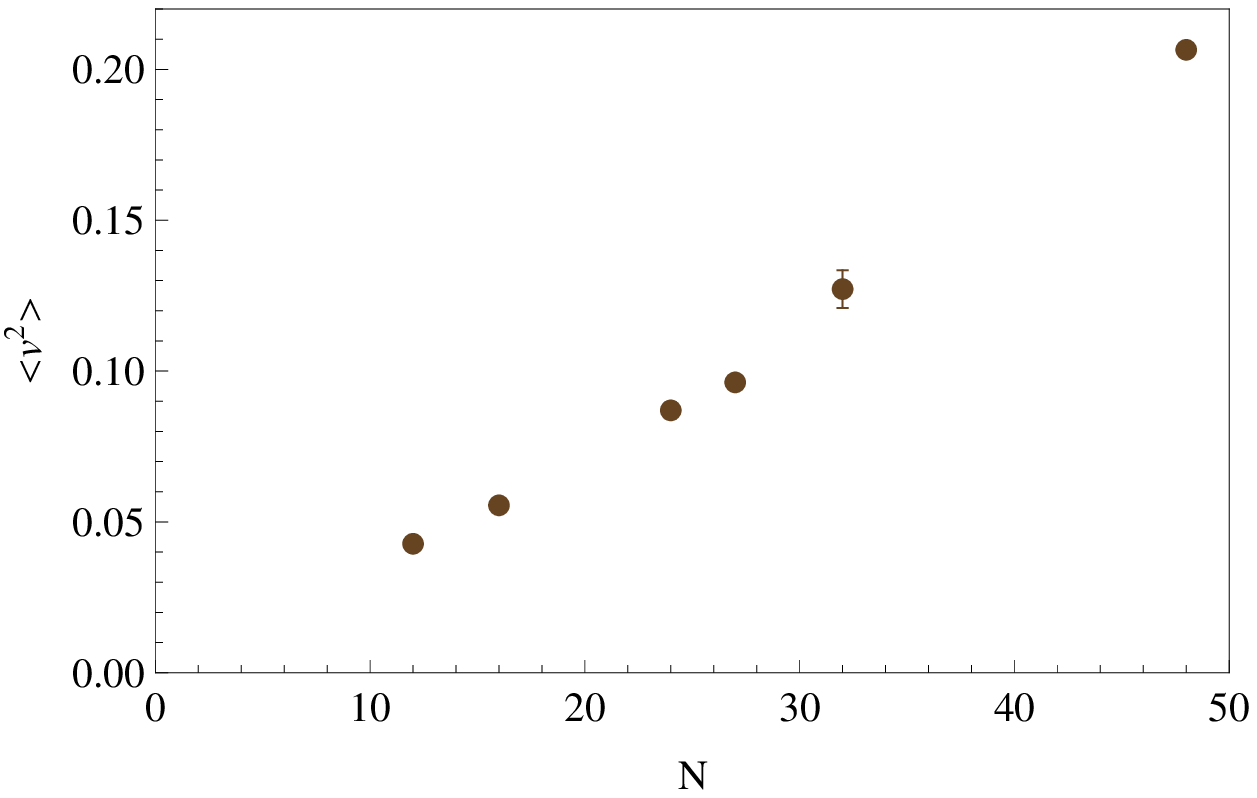}
\caption{\label{KEvsN} The kinetic energy as a function of the $N$, the number of particles in the simulation for the cases without adaptivity (Fig.~\ref{fig:Epsilon}).}
\end{figure}

\subsection{The simulation with softened gravity (``epsilon-gravity'')}

\begin{figure}
\centering
\includegraphics[width=0.7\linewidth]{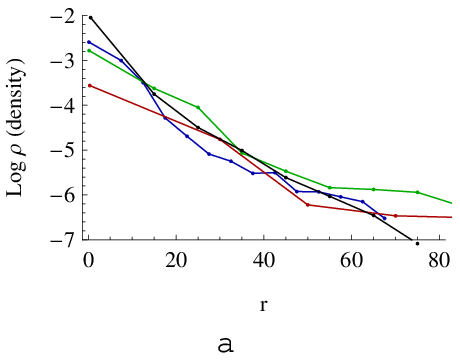}
\includegraphics[width=\linewidth]{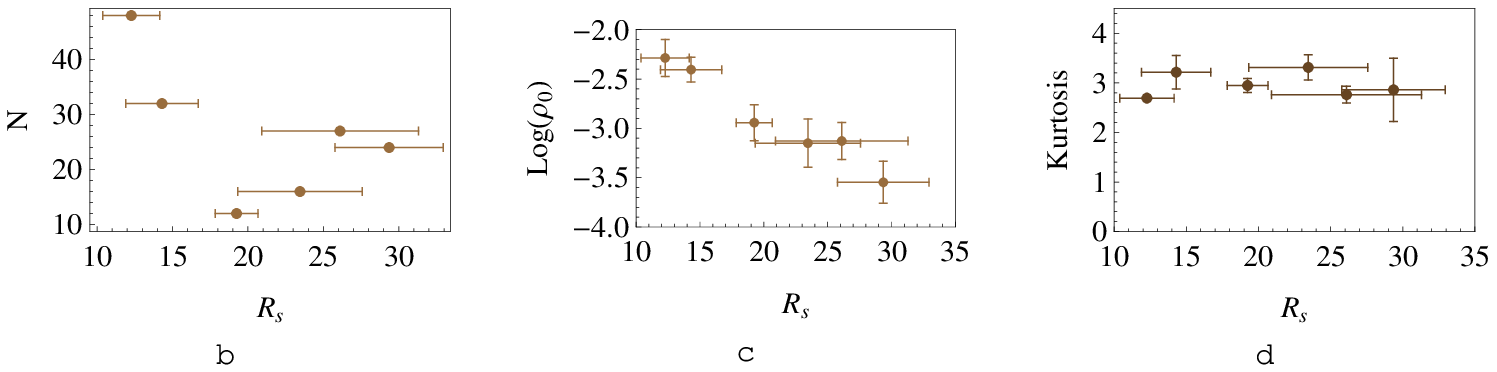}
\caption{\label{fig:Epsilon}
(a) - Density profiles of simulated clusters with the softened-gravity (``epsilon gravity''). In Black $N=48$, $R_s=9.29$. Green - $N=27$, $R_s=26.08$. Red - $N=16$, $R_s=30$. Blue - $N=12$, $R_s=16.4$.
(b),(c),(d) - The total number of midges, the density at the center and the kurtosis vs. the size of the swarm for the swarms in the ``epsilon gravity''. ($\epsilon^2=15$)}
\end{figure}

In order to avoid high accelerations due to close encounters (``slingshots'') and thus evaporation of the cluster we included a softening parameter $\epsilon$ to the gravitational force (Eq.~1).
The particles in the simulation still developed high accelerations in close encounters relative to simulations with adaptivity since adaptivity itself contributed to the decrease of accelerations by $\epsilon^2/(R_{ad}^2+\epsilon^2)$, and in order to compare the simulations in the two cases (with and without adaptivity) we have to reduce the accelerations in the ``$\epsilon$-gravity'' case by this constant factor. The results in Fig.~\ref{fig:Epsilon} were obtained this way. In Fig.~\ref{fig:Epsilon}(a) density profiles for various sizes of clusters are given (where $\epsilon^2=15$) and in Fig.~\ref{fig:Epsilon}(b)-(d) we show three characteristic quantities of the clusters: the number of particles, the density at the center and the kurtosis. Here and in all the paper we truncated the tails at $r \sim 3\,R_s$ in the calculation of the kurtosis.

\vspace{10 mm}

\bibliography{acoustics_1}